\documentstyle[preprint,prc,aps,psfig]{revtex}

\renewcommand{\vec}[1]{\mbox{\boldmath $#1$}}

\tightenlines

\begin{document}

\title{Large-angle scattering and quasi-elastic barrier distributions} 
\author{K. Hagino$^{1}$ and N. Rowley$^2$}
\address{$^1$Yukawa Institute for Theoretical Physics, Kyoto
University, Kyoto 606-8502, Japan }
\address{$^2$Institut de Recherches Subatomiques, 
UMR7500, IN2P3-CNRS/Universtit\'e Louis Pasteur, BP28, F-67037 
Strasbourg Cedex 2, France}

\maketitle

\begin{abstract}

We study in detail the barrier distributions extracted from 
large-angle quasi-elastic scattering of heavy ions at energies
near the Coulomb barrier. Using a closed-form expression for scattering 
from a single barrier, we compare the quasi-elastic barrier 
distribution with the corresponding test function for fusion.
We examine the isocentrifugal approximation 
in coupled-channels calculations of quasi-elastic scattering
and find that for backward angles, it works well, 
justifying the concept of a barrier distribution 
for scattering processes. This method
offers an interesting tool for investigating unstable nuclei.
We illustrate this for the $^{32}$Mg + $^{208}$Pb reaction, 
where the quadrupole collectivity of the neutron-rich $^{32}$Mg 
remains to be clarified experimentally. 

\end{abstract}


\section{Introduction}

Heavy-ion collisions at energies around the Coulomb barrier 
provide an ideal opportunity to study quantum tunneling phenomena 
in systems with many degrees of freedom \cite{DHRS98,BT98}. 
In a simple model, a potential barrier 
for the relative 
motion between the colliding nuclei is created by the strong
interplay of the repulsive Coulomb force with the attractive
nuclear interaction. In the eigenchannel approximation, this barrier
is split into a number of distributed barriers due to couplings of the
relative motion to intrinsic degrees of freedom (such as collective
inelastic excitations of the colliding nuclei and/or transfer processes), 
resulting in the subbarrier enhancement of fusion cross sections \cite{DLW83}. 
It is now well known that a barrier
distribution can be extracted experimentally 
from the fusion excitation function 
$\sigma_{\rm fus}(E)$ by taking the second derivative of the product
$E\sigma_{\rm fus}(E)$ with respect to the center-of-mass energy $E$, that is, 
$d^2(E\sigma_{\rm fus})/dE^2$. This method was first proposed by Rowley,
Satchler, and Stelson in
Ref. \cite{RSS91}, and has stimulated precise measurements of fusion
cross sections for many systems \cite{L95,SACN95} (see Ref. \cite{DHRS98}
for a detailed review). The extracted fusion barrier distributions have been
found to be very sensitive to the structure of the colliding nuclei,
and thus the barrier distribution method has opened up 
the possibility of 
using the heavy-ion fusion reaction as a ``quantum tunneling
microscope'' in order to investigate both the static and 
dynamical properties of atomic nuclei. 

Channel couplings also affect the scattering process. 
In Ref. \cite{ARN88}, it 
was suggested 
that the same information as the fusion cross
section may be obtained from the cross section for quasi-elastic
scattering (a sum of elastic, inelastic, and transfer cross sections) 
at large angles. 
At these backward angles, it is known that the
single-barrier elastic cross section falls off smoothly from a value
close to that for Rutherford scattering at low energies to very small
values at energies high above the barrier. Timmers {\it et al.}
therefore proposed to use the first derivative of the ratio of the
quasi-elastic cross section $\sigma_{\rm qel}$ 
to the Rutherford cross section $\sigma_R$ with
respect to energy, $-d (d\sigma_{\rm qel}/d\sigma_R)/dE$, as an
alternative 
representation of the barrier distribution \cite{TLD95}. The
experimental data of Timmers {\it et al.} have revealed \cite{TLD95} 
that the quasi-elastic barrier distribution is indeed similar to that
for fusion, although the former may be somewhat smeared and thus 
less sensitive to nuclear structure effects. 

There are certain attractive experimental advantages to 
measuring the quasi-elastic 
cross section $\sigma_{\rm qel}$ rather than the fusion cross sections 
$\sigma_{\rm fus}$ to extract a representation of 
the barrier distribution\cite{R03}. 
These are: i) less accuracy is required in the data for taking the
first derivative rather than the second derivative, ii) whereas
measuring the fusion cross section requires specialized recoil
separators (electrostatic deflector/velocity filter) usually of low
acceptance and efficiency, the measurement of 
$\sigma_{\rm qel}$ needs only very simple charged-particle detectors,
not necessarily possessing good resolution either in energy or in
charge, and iii) several effective energies can be measured at a
single-beam energy, since, in the semi-classical approximation, each 
scattering angle corresponds to scattering at a certain angular
momentum, and the cross section can be scaled in energy by taking 
into account the centrifugal correction. 
The last point not only improves the efficiency of the experiment, 
but also allows the use of a cyclotron accelerator where the
relatively small energy steps required for barrier 
distribution experiments cannot be obtained from the
machine itself. This fact was recently exploited by Piasecki 
{\it et al.}\cite{PKP02}, 
who took an astute choice of the scattering angles at which 
$\sigma_{\rm qel}$ was measured in order to have the energy range
necessary, while retaining relatively small energy steps. 
Moreover, these advantages all point to greater ease of measurement with
low-intensity exotic beams. 

In this paper, we 
undertake a detailed discussion of the
properties of the quasi-elastic barrier distribution. 
In contrast to the fusion barrier distribution, a theoretical
description of the quasi-elastic barrier distribution has been 
limited so far either to a purely classical level or to a completely
numerical level. 
Given that many new barrier distribution 
measurements for exotic nuclei 
are expected to come out in the near future, due to an increasing
availability of radioactive beams, we believe that 
it is of considerable importance to clarify 
the properties of the quasi-elastic barrier distribution 
in a more reliable and transparent way. 

The paper is organized as follows. 
In Sec. II, we consider a single-barrier system and discuss test
functions for the barrier distribution, that is the representations of
the barrier distribution for a single barrier case. 
We first briefly review the fusion test function, and discuss 
the relation to the barrier penetrability. We then use 
semi-classical perturbation theory \cite{B85,LW81} 
to derive an analytical expression
for the elastic cross section at backward angles. 
Using the formula thus obtained, we discuss the energy dependence of
the quasi-elastic test function, and compare it with that for
the fusion test function. We also discuss the scaling property
of the quasi-elastic test function obtained at different scattering
angles. 
In Sec. III, we discuss the barrier distribution for coupled-channels
systems. Theoretically, barrier
distributions have a clear physical meaning 
only in the limit of zero angular momentum transfer 
(that is, in the isocentrifugal approximation
\cite{LR84,NRL86,NBT86,HTBB95,ELP87,T87,TMBR91,AA94,GAN94}) 
with vanishing excitation 
energies for the intrinsic degrees of freedom. In this limit the 
barrier distribution representation may be derived analytically as a 
weighted sum of test functions.
Nevertheless, a simple two-level model suggests that the concept 
holds to a good approximation even when the excitation
energy is finite \cite{HTB97}. And of course many experimental
data also show well-defined barrier structures, which can be reproduced
by coupled-channels calculations, even for systems where 
the excitation energies are large.
However, although the validity of the 
isocentrifugal approximation has been shown to work well for
fusion \cite{T87}, its applicability for scattering processes
in the presence of the
long-range Coulomb interaction is less clear
\cite{ELP87,TMBR91,AA94,GAN94}. 
We therefore re-examine its validity for the quasi-elastic barrier
distribution. 
In Sec. IV, we consider the quasi-elastic barrier distribution 
as applied to reactions induced by exotic nuclei. In particular,
we demonstrate its usefulness by showing the possible effects of the
quadrupole excitation of $^{32}$Mg in the $^{32}$Mg + $^{208}$Pb system. 
We summarize the paper in Sec. V. 

\section{Single-barrier problems}

In this section, we discuss heavy-ion reactions between inert nuclei. 
For such a system, the incident flux of the projectile is either 
absorbed or elastically scattered from the target nucleus. 
We use a local optical potential which is energy and angular momentum 
independent. Assuming that the imaginary part of the optical
potential is strong enough and is localized well inside the Coulomb
barrier, the absorption cross section is identified with the
fusion cross section. 

\subsection{Fusion test function}

Let us first discuss the properties of the fusion test function. 
The classical fusion cross section is given by, 
\begin{equation}
\sigma^{cl}_{\rm fus}(E)=\pi
R_b^2\left(1-\frac{B}{E}\right)\,\theta(E-B),
\end{equation}
where $R_b$ and $B$ are the barrier position and the barrier height,
respectively. 
From this expression, it is clear that the first derivative of 
$E\sigma^{cl}_{\rm fus}$ is proportional to the classical 
penetrability for a 1-dimensional barrier of height $B$ or
eqivalently the s-wave penetrability, 
\begin{equation}
\frac{d}{dE}[E\sigma^{cl}_{\rm fus}(E)]=\pi R_b^2\,\theta(E-B)
=\pi R_b^2\,P_{cl}(E),
\end{equation}
and the second derivative to a delta function, 
\begin{equation}
\frac{d^2}{dE^2}[E\sigma^{cl}_{\rm fus}(E)]=\pi R_b^2\,\delta(E-B). 
\label{clfus}
\end{equation}

In quantum mechanics, 
the tunneling effect smears the delta function in Eq. (\ref{clfus}). 
An analytic formula for  
the fusion cross section can be obtained if one approximates 
the Coulomb barrier as an inverse parabola, and is given by \cite{W73}, 
\begin{equation}
\sigma_{\rm fus}(E)=\frac{\hbar\Omega}{2E}
R_b^2\ln\left[1+e^{2\pi(E-B)/\hbar\Omega}\right], 
\end{equation}
where $\hbar\Omega$ is the curvature of the Coulomb barrier. 
Again, the 
first derivative of 
$E\sigma_{\rm fus}$ is proportional to the s-wave penetrability for a
parabolic barrier, 
\begin{equation}
\frac{d}{dE}[E\sigma_{\rm fus}(E)]=\pi R_b^2\,
\frac{1}{1+\exp\left[-\frac{2\pi}{\hbar\Omega}(E-B)\right]} 
=\pi R_b^2\,P(E). 
\label{Ppara}
\end{equation}
Defining the function $G_{\rm fus}(E)$ as, 
\begin{equation}
G_{\rm fus}(E)\equiv\frac{1}{\pi R_b^2}\,\frac{d^2}{dE^2}
[E\sigma_{\rm fus}(E)], 
\end{equation}
eq. (\ref{Ppara}) leads to 
\begin{equation}
G_{\rm fus}(E)=\frac{dP(E)}{dE} 
=\frac{2\pi}{\hbar\Omega}\,\frac{e^x}{(1+e^x)^2},
\label{Gfus}
\end{equation}
where $x\equiv 2\pi(E-B)/\hbar\Omega$. This function has the
following properties: 
i) it is symmetric around $E=B$, ii) it is centered on $E=B$, iii) 
its integral over $E$ is unity, and iv) it has a relatively narrow width
of around $\hbar\Omega\ln(3+\sqrt{8})/\pi \sim 0.56 \hbar\Omega$. 
In the next section, we will show that a barrier distribution can be 
expressed as a weighted sum of normalized functions $G(E)$ 
(see Eq. (\ref{weightedsum})). 
The function 
$G_{\rm fus}(E)$, therefore, plays the role of a test function, and we
call it the fusion test function. 

\subsection{Quasi-elastic test function}

We now ask ourselves 
the question of how best to define a similar test function 
for a scattering problem. 
In the pure classical approach, in the limit of a strong Coulomb field, 
the differential cross sections for elastic scattering at $\theta=\pi$ 
is given by, 
\begin{equation}
\sigma_{\rm el}^{cl}(E,\pi)=\sigma_R(E,\pi)\,\theta(B-E), 
\end{equation}
where $\sigma_R(E,\pi)$ is the Rutherford cross section. 
Thus, the ratio $\sigma_{\rm el}^{cl}(E,\pi)/\sigma_R(E,\pi)$ is the
classical reflection probability $R(E)$ ($=1-P(E)$), and 
Eq. (\ref{Gfus}) suggests that the appropriate test function for 
scattering is \cite{TLD95},
\begin{equation}
G_{\rm qel}(E)=-\frac{dR(E)}{dE} 
=-\frac{d}{dE}\left(\frac{\sigma_{\rm el}(E,\pi)}{\sigma_R(E,\pi)}\right). 
\label{qeltest}
\end{equation}

In realistic systems, however, due to the effect of nuclear
distortion, the differential cross section deviates from the
Rutherford cross section even at energies below the barrier. 
Using the semi-classical perturbation theory \cite{B85,LW81,BS81}, 
we derive in the Appendix 
a semi-classical formula for the backward scattering 
which takes into account the nuclear effect to the leading order. 
The result for a scattering angle $\theta$ 
reads, 
\begin{equation}
\frac{\sigma_{\rm el}(E,\theta)}{\sigma_R(E,\theta)}
=\alpha(E,\lambda_c)\cdot |S(E,\lambda_c)|^2,
\label{ratio}
\end{equation}
where 
$S(E,\lambda_c)$ is the total (Coulomb + nuclear)
$S$-matrix at energy $E$ and angular momentum 
$\lambda_c = \eta\cot(\theta/2)$, with $\eta$ being the usual Sommerfeld
parameter. 
Note that 
$|S(E,\lambda_c)|^2$ is nothing but the reflection probability of the
Coulomb barrier. For $\theta=\pi$, $\lambda_c$ is zero, and 
$|S(E,\lambda_c=0)|^2$ is given by 
\begin{equation}
|S(E,\lambda_c=0)|^2 = R(E) = 
\frac{\exp\left[-\frac{2\pi}{\hbar\Omega}(E-B)\right]}
{1+\exp\left[-\frac{2\pi}{\hbar\Omega}(E-B)\right]} 
\end{equation}
in the parabolic approximation. 
$\alpha(E,\lambda_c)$ in Eq. (\ref{ratio}) is given by 
\begin{equation}
\alpha(E,\lambda_c)=1+\frac{V_N(r_c)}{ka}\,
\frac{\sqrt{2a\pi k\eta}}{E}\,
\left[1-\frac{r_c}{Z_PZ_Te^2}\cdot
2V_N(r_c)\left(\frac{r_c}{a}-1\right)\right],
\end{equation}
where $k=\sqrt{2\mu E/\hbar^2}$, with $\mu$ being the reduced mass for
the colliding system. The nuclear potential $V_N(r_c)$ is evaluated at
the Coulomb turning point $r_c=(\eta+\sqrt{\eta^2+\lambda_c^2})/k$,
and $a$ is the diffuseness parameter in the nuclear potential. 

The upper panel of Fig. 1 shows the excitation function of the
cross sections
at $\theta=\pi$ for the $^{16}$O + $^{144}$Sm reaction. We use 
an optical potential of the Woods-Saxon form, with parameters 
$V_0$ = 105.1 MeV, $r_0$ = 1.1 fm, $a$ = 0.75 fm, 
$W$=30 MeV, $r_W$ = 1.0 fm, and $a_W$ = 0.4 fm. 
The solid line is the exact solution of the Schr\"odinger equation,
while the dashed line is obtained with the semi-classical formula
(\ref{ratio}). The dotted line shows the reflection probability
$R(E)=|S(E)|^2$. 
We clearly see that the semi-classical formula accounts well for
the deviation of the elastic cross section $\sigma_{\rm el}(E)$ from
the Rutherford cross section around the Coulomb barrier. 

The corresponding quasi-elastic test functions,  
$G_{\rm qel}(E)=-d/dE(\sigma_{\rm el}/\sigma_R)$, are 
shown in the lower panel of Fig. 1. 
We use a point-difference formula with $\Delta E_{\rm cm}=1.8$ MeV 
(as in an experiment) in order to
evaluate the energy derivative. 
Notice that the first derivative of 
the reflection probability (dotted line) corresponds to the fusion 
test function $G_{\rm fus}(E)$ given in Eq. (\ref{Gfus}). 
Because of the nuclear distortion factor $\alpha(E,\lambda_c)$, 
the quasi-elastic test function behaves a little less simply than that for
fusion. We find: i) the peak position slightly deviates from the 
barrier height $B$ (by 0.265 MeV for the example shown in Fig.~1), 
and ii) it has a low-energy tail. 
Eq. (\ref{ratio}) indicates that there are two contributions to 
the quasi-elastic test function. One is $\alpha(E)\cdot dR(E)/dE$, and the 
other $d\alpha(E)/dE\cdot R(E)$. 
In Fig.2, we show these two contributions separately. 
We notice that the low-energy tail of the quasi-elastic test function 
comes from the latter, that is, the energy dependence of the nuclear 
distortion factor $\alpha(E)$. 

Despite these small drawbacks, the quasi-elastic test function 
$G_{\rm qel}(E)$ behaves rather similarly to the fusion test function 
$G_{\rm fus}(E)$. 
In particular, 
both functions have a similar, relatively narrow, width, 
and their integral over $E$ is unity. 
We may thus consider that the quasi-elastic test function is an excellent 
analogue of the one for fusion, and we exploit this fact in 
studying barrier structures in heavy-ion scattering. 

Notwithstanding the above comments, it is clear that the 
quasi-elastic test function defined above depends on 
the scattering angle, and below we shall show how the test
function can be scaled in terms of an effective energy.  

\subsection{Scaling property of the quasi-elastic test function}

One of the advantages of the quasi-elastic test function over the 
fusion test function is that different scattering angles correspond to 
the different grazing angular momenta. To some extent, the effect of 
angular momentum can be corrected by shifting the energy by an amount 
equal to the centrifugal 
potential. Estimating the centrifugal potential at the Coulomb turning 
point $r_c$, the effective energy may be expressed as \cite{TLD95}
\begin{equation}
E_{\rm eff}\sim E
-\frac{\lambda_c^2\hbar^2}{2\mu r_c^2} 
=2E\frac{\sin(\theta/2)}{1+\sin(\theta/2)}.
\label{Eeff}
\end{equation}
In deriving this equation, we have used the definition of $r_c$, that is, 
$E=Z_PZ_Te^2/r_c+\lambda_c^2\hbar^2/2\mu r_c^2$. 
Therefore, one expects that the function 
$-d/dE (\sigma_{\rm el}/\sigma_R)$ 
evaluated 
at an angle $\theta$ will correspond to the quasi-elastic test function 
(\ref{qeltest}) 
at the effective energy given by eq. (\ref{Eeff}). 

In order to check the scaling property of the quasi-elastic test function 
with respect to the angular momentum, 
Fig. 3 compares the functions  
$\sigma_{\rm el}/\sigma_R$ (upper panel) and 
$-d/dE (\sigma_{\rm el}/\sigma_R)$ (lower panel) 
obtained at two different scattering angles. 
The solid line is evaluated at $\theta=\pi$, while the dotted line at 
$\theta=160^{\rm o}$. The dashed line is the same as the dotted line, but 
shifted in energy by $E_{\rm eff}-E$. 
Evidently, the scaling does work well, both at 
energies below and above the Coulomb barrier. 

We should note, however, that as the scattering angle decreases,
the scaling becomes less good. 
See Fig. 4 for the scaling property for 
$\theta=140^{\rm o}$. 
Thus in planning 
an experiment (especially if it combines data taken in detectors 
at different 
angles), one should take careful account of this effect. Also at
smaller angles, it is well known that the underlying elastic cross
section will display Fresnel oscillations, which would cause the test
function itself (and any derived distribution) to oscillate. 
Detector angles are best chosen to minimise effects of Fresnel
oscillations. 

\section{Barrier distribution for multi-channel systems}

\subsection{Barrier distributions in the sudden tunneling limit}

Let us now discuss the barrier distributions in the presence of 
a coupling between the relative motion $\vec{r}$ 
and an intrinsic
degree of freedom $\xi$. The standard way to address the effect of the
coupling is to solve the coupled-channels equations. 
For a problem of heavy-ion fusion reactions, these 
equations are often solved in the iso-centrifugal approximation 
\cite{HRK99}, 
where 
one replaces the angular momentum of the relative motion in each
channel by the total angular momentum $J$ 
(this 
approximation is also referred to as the rotating frame approximation 
or the no-Coriolis approximation in the literature). 
The iso-centrifugal approximation dramatically simplifies the angular
momentum couplings, and reduces the dimension
of the coupled-channels equations in a considerable way 
\cite{LR84,NRL86,NBT86,HTBB95,ELP87,T87,TMBR91,AA94,GAN94}. 
The coupled-channels equations in this approximation are given by 
\begin{eqnarray}
\left(-\frac{\hbar^2}{2\mu}\frac{d^2}{dr^2}+ 
\frac{J(J+1)\hbar^2}{2\mu r^2}+V_0(r)-E+\epsilon_I\right)
u_{I}(r)&& \nonumber \\
+\sum_{I'}
\sqrt{\frac{2\lambda+1}{4\pi}}f(r)
\langle\varphi_{I0}|T_{\lambda 0}|\varphi_{I'0}\rangle 
u_{I'}(r)&=&0, 
\label{ccisocent}
\end{eqnarray}
where $|\varphi_{IM}\rangle$ is an intrinsic wave function which
satisfies $H_{\rm int}|\varphi_{IM}\rangle =
\epsilon_I|\varphi_{IM}\rangle$. 
We have assumed that 
the coupling Hamiltonian 
is given by 
$V_{\rm coup}=f(r)Y_{\lambda\mu}(\hat{\vec{r}})\hat{T}^*_{\lambda\mu}(\xi)$. 
The coupled-channels equations are solved with the scattering boundary 
condition for $u_I(r)$,
\begin{equation}
u_{I}(r)\to\frac{i}{2}\left\{ H_J^{(-)}(k_ir)\delta_{I,I_i}
-\sqrt{\frac{k_i}{k_{I}}}S^J_{I}
H_{J}^{(+)}(k_{I}r)
\right\}, 
\end{equation}
where $S^J_{I}$ is the nuclear S-matrix. $H_{l}^{(-)}(kr)$ and 
$H_{l}^{(+)}(kr)$ are the incoming and the outgoing Coulomb wave
functions, respectively. The channel wave number $k_{I}$ is 
given by $\sqrt{2\mu(E-\epsilon_I)/\hbar^2}$, and
$k_i=k_{I_i}=\sqrt{2\mu E/\hbar^2}$. 
The scattering angular distribution for the channel $I$ is then 
given by \cite{ELP87}, 
\begin{equation}
\frac{d\sigma_I}{d\Omega}
=\frac{k_I}{k_i}|f_{I}(\theta)|^2, 
\end{equation}
with 
\begin{equation}
f_I(\theta)=\sum_J
e^{i[\sigma_{J}(E)+\sigma_J(E-\epsilon_I)]} 
\sqrt{\frac{2J+1}{4\pi}}\,Y_{J0}(\theta)\,
\frac{-2i\pi}{\sqrt{k_ik_{lI}}}(S^J_I-\delta_{I,I_i})
+f_C(\theta)\delta_{I,I_i}, 
\end{equation}
where $\sigma_J(E)$ and $f_C(\theta)$ are the Coulomb phase shift and 
the Coulomb scattering amplitude, respectively. 

In the limit of $\epsilon_I\to 0$, the reduced coupled-channels
equations (\ref{ccisocent}) are completely decoupled. In this limit,
the coupling matrix defined as 
\begin{equation}
V_{II'}\equiv \epsilon_I\delta_{I,I'} 
+\sqrt{\frac{2\lambda+1}{4\pi}}f(r)
\langle\varphi_{I0}|T_{\lambda 0}|\varphi_{I'0}\rangle 
\label{coup}
\end{equation}
can be diagonalized independently of $r$. 
It is then easy to prove that the fusion and the quasi-elastic 
cross sections are given as a weighted sum of the cross sections 
for uncoupled eigenchannels\cite{NRL86,NBT86}, 
\begin{eqnarray}
\sigma_{\rm fus}(E)&=&\sum_\alpha w_\alpha 
\sigma_{\rm fus}^{(\alpha)}(E), \label{crossfus}\\
\sigma_{\rm qel}(E,\theta)&=&\sum_I\sigma_I(E)=\sum_\alpha w_\alpha 
\sigma_{\rm el}^{(\alpha)}(E,\theta), \label{crossqel}
\end{eqnarray}
where $\sigma_{\rm fus}^{(\alpha)}(E)$ and 
$\sigma_{\rm el}^{(\alpha)}(E,\theta)$
are the fusion and the elastic cross sections for a potential 
in the eigenchannel $\alpha$, that is, 
$V_\alpha(r)=V_0(r)+\lambda_\alpha(r)$. Here, 
$\lambda_\alpha(r)$ is the eigenvalue of the coupling matrix
(\ref{coup}) (when $\epsilon_I$ is zero, $\lambda_\alpha(r)$ is 
simply given by $\lambda_\alpha\cdot f(r)$). 
The weight factor $w_\alpha$ is given by 
$w_\alpha=U_{0\alpha}^2$, where $U$ is the unitary matrix which diagonalizes
Eq. (\ref{coup}). 
Eqs. (\ref{crossfus}) and (\ref{crossqel}) immediately lead to 
the expressions for the barrier distribution in terms of the test
functions introduced in the previous section, 
\begin{eqnarray}
D_{\rm fus}(E)&=&\frac{d^2}{dE^2}[E\sigma_{\rm fus}(E)]=
\sum_\alpha w_\alpha 
\pi R_{b,\alpha}^2\,G_{\rm fus}^{(\alpha)}(E), 
\label{weightedsum}
\\
D_{\rm qel}(E)&=&
-\frac{d}{dE}\left(\frac{\sigma_{\rm qel}(E,\pi)}{\sigma_R(E,\pi)}\right) 
=
\sum_\alpha w_\alpha 
G_{\rm qel}^{(\alpha)}(E). 
\end{eqnarray}

As an example of these formulas, let us consider the effect of
rotational excitations of the target nucleus 
in the reaction of $^{16}$O with the 
deformed $^{154}$Sm.  For this problem, cross 
sections (\ref{crossfus}) and
(\ref{crossqel}) can be computed as \cite{RHT01}
\begin{equation}
\sigma(E)=\int^1_0d(\cos\theta_T)\sigma(E;\theta_T),
\label{orientation}
\end{equation}
where $\theta_T$ is the orientation of the deformed target. The 
angle dependent potential $V(r,\theta_T)$ is given by,
\begin{eqnarray}
V(r,\theta_T)&=&V_N(r,\theta_T)+V_C(r,\theta_T), \\
V_N(r,\theta_T)&=&\frac{-V_0}{1+\exp[(r-R-R_T\beta_2Y_{20}(\theta_T)
-R_T\beta_4Y_{40}(\theta_T))]}, \\
V_C(r,\theta_T)&=&\frac{Z_PZ_Te^2}{r}+\sum_\lambda\left(\beta_\lambda
+\frac{2}{7}\sqrt{\frac{5}{\pi}}\beta_2^2\delta_{\lambda,2}\right)
\,\frac{3Z_PZ_Te^2}{2\lambda+1}\frac{R_T^\lambda}{r^{\lambda+1}}
Y_{\lambda0}(\theta_T). 
\end{eqnarray}
Figs. 5(a) and 5(b) 
show the barrier distributions obtained with Eq. (\ref{orientation}) 
for the fusion and the quasi-elastic processes, respectively. 
We use the potential whose parameters are 
$V_0$=220 MeV, $R=1.1\times (A_T^{1/3}
+A_P^{1/3})$ fm, and $a$=0.65 fm. The deformation parameters 
are taken to be $\beta_2$=0.306 and $\beta_4$=0.05. 
We replace the integral in Eq. (\ref{orientation}) with the 
$(I_{\rm max}+2)$-point Gauss quadrature \cite{NBT86} with 
$I_{\rm max}$=10. That is, we take 6 different 
orientation angles. The contributions from each eigenbarrier 
are shown by the dashed line in Fig. 5(a) and Fig. 5(b). 
The solid line is the sum of all the contributions, which is compared 
with the experimental data \cite{L95,TLD95}. The agreement between 
the calculation and the experimental data is
reasonable both for the fusion and the quasi-elastic barrier
distributions. For the fusion barrier distribution $D_{\rm fus}$, 
the agreement will be further improved if one uses a larger value of
diffuseness parameter $a$ \cite{L95,HRD03} (see the dotted line). 
Fig. 5(c) compares the fusion with the quasi-elastic barrier
distributions. These are normalized so that the energy integral
between 50 and 70 MeV is unity. 
As we discussed in Sec. II for a single barrier case, we see 
that the two barrier distributions show a very similar behavior 
to each other. 

\subsection{Barrier distributions in systems with finite excitation
energy}

In general, the approximation of neglecting the excitation energies
$\epsilon_I$ (that is, the sudden tunneling approximation) 
is valid only for rotational states in heavy deformed nuclei. 
Despite this, however, some of the most interesting effects have been 
found in the fusion barrier distributions for 
systems involved with highly vibrational nuclei as well \cite{L95,SACN95}. 
One finds that the barrier structures still exist, but that the
weights of the different barriers can be strongly influenced by 
non-adiabatic effects. 
In Ref. \cite{HTB97}, we have explicitly demonstrated that the fusion 
cross sections are in general given by Eq. (\ref{crossfus}), but 
with the energy
dependent weight factors $w_\alpha(E)$ (in the sudden tunneling limit,
the weight factors become energy independent). 
For a simple two-channel problem, we found that although the weights 
may depend strongly on the excitation energy, their dependence on the
incident energy is weak, suggesting 
that the concept of a barrier distribution holds good
even for finite intrinsic excitation energies \cite{HTB97}.
Since the quasi-elastic barrier distribution $G_{\rm qel}(E)$ is
related to the fusion barrier distribution $G_{\rm fus}(E)$ through
flux conservation (unitarity), a similar situation can be expected for the 
quasi-elastic barrier distribution. 

\subsection{Applicability of the iso-centrifugal approximation}

As we have mentioned in Sec. I, the validity of the iso-centrifugal
approximation has been well tested for heavy-ion fusion
reactions\cite{T87}. In contrast, it is known that the approximation 
fails to reproduce the exact result for scattering angular
distributions in the presence of the long-range 
Coulomb force. 
The effect of the coupling is somewhat overestimated in the
isocentrifugal approximation, and simple recipes to renormalize the
coupling strength have been proposed in order to cure this problem 
\cite{ELP87,TMBR91,AA94,GAN94}. 
On the other hand, Esbensen {\it et al.} have argued, based on
semi-classical considerations, that the 
isocentrifugal approximation (without renormalization of the 
coupling strength) works better for backward angle
scattering \cite{ELP87}. 

Since it has not yet been clear how well the isocentrifugal
approximation works in connection with the quasi-elastic barrier
distribution, we re-examine in this subsection the performance of the 
approximation for large-angle scattering. 
To this end, we consider the effect of quadrupole phonon excitations
in the target nucleus for the $^{16}$O + $^{144}$Sm reaction. 
In order to emphasise the coupling effect, we increase the coupling
strength and reduce the excitation energy from the physical values. The 
values which we use are: 
$\beta_2$=0.2 (with $r_{\rm coup}$ = 1.06 fm) and $\epsilon_2$=0.5 MeV. 
We have checked that our
conclusions are not altered irrespective of the values of 
$\beta_2$ and $\epsilon_2$. 
For simplicity, we consider only a single phonon excitation,
and employ the linear coupling approximation \cite{HTDHL97}. 
We use the same optical potential as in Sec. II. 

Figure 6 shows the partial cross sections at $E_{\rm cm}$=65 MeV 
for the angle-integrated 
inelastic scattering (upper panel) and for the fusion reaction
(lower panel) as a function of the initial orbital angular
momentum $l_i=J$. 
The solid line is the exact result of the coupled-channels equations
with the full angular momentum couplings, while the dashed line is 
obtained with the iso-centrifugal approximation. 
We find that the isocentrifugal approximation works rather well for 
$J \leq 20$, although the agreement is poor for larger values of $J$. 
For fusion, only small values of $J$ contribute, and the
isocentrifugal approximation always makes an excellent approximation. 
Fig. 7 shows the angular distributions for the elastic
(upper panel) and inelastic scattering (lower panel). 
Although the isocentrifugal approximation does not reproduce the main
structure of the angular distribution, it indeed works very nicely at 
backward angles where 
the main contribution comes from 
small values of angular momentum (see Eq. (\ref{ratio}) and Fig. 6). 
In fact, the isocentrifugal approximation almost reproduces 
the exact result for the scattering angles $\theta_{cm}>130^{\rm o}$.

Fig. 8 shows the excitation function for quasi-elastic scattering
(upper panel) and its energy derivative calculated at 
$\theta=170^{\rm o}$ in the laboratory frame. One sees that the
isocentrifugal approximation well reproduces the exact solution. 
We thus conclude that the 
the isocentrifugal approximation works sufficiently well for 
studies of quasi-elastic barrier distributions. 
This fact not only makes the coupled-channels calculations
considerably easier, but also assures the similarity of fusion
and quasi-elastic distributions even in the presence of
channel couplings. 

\section{Quasi-elastic scattering with radioactive beams}

It has been well recognized that low-energy reactions 
provide an ideal tool to probe the detailed structure of atomic
nuclei. The heavy-ion fusion reaction around the Coulomb barrier is 
one of the typical examples. In the last decade, many high-precision
measurements of fusion cross sections have been made, and the nuclear 
structure information has been successfully extracted through the
representation of the fusion barrier
distribution \cite{DHRS98}. 

Low-energy radioactive beams have also become increasingly 
available in recent years,
and heavy-ion fusion reactions involving neutron-rich nuclei 
have been performed for a few systems \cite{S01,SYW04,YSF96,TSA00,LSG03}. 
New generation facilities have been under 
construction at several laboratories, and many more 
reaction measurements 
with exotic beams at low energies will be performed in the near future
(see Ref. \cite{AN03} for a recent theoretical review). 
Although it would still be difficult to perform high-precision
measurements of fusion cross sections with radioactive beams, 
the measurement of the quasi-elastic barrier distribution, which 
can be obtained much more easily than the fusion
counterpart as we mentioned in the introduction, may be feasible.
Since the quasi-elastic barrier distribution contains similar
information as the fusion barrier distribution, 
the quasi-elastic measurements at backward angles 
may open up a
novel way to probe the structure of exotic neutron-rich nuclei. 

In order to demonstrate the usefulness of the study of the 
quasi-elastic barrier
distribution with radioactive beams, we take as an example the
reaction $^{32}$Mg and $^{208}$Pb. 
The neutron-rich $^{32}$Mg nucleus has attracted much interest as 
evidence for the breaking of the $N$=20 spherical shell closure. 
In this nucleus, a large $B(E2)$ value (454$\pm$78 $e^2$fm$^4$
\cite{MIA95} and 622$\pm$ 90 $e^2$fm$^4$ \cite{CGL01}) and a small 
value of the excitation energy of the first 2$^+$ state (885 keV)
\cite{MIA95} have been experimentally observed. 
The authors of Refs. \cite{MIA95,CGL01,YSG01} argue that these large
collectivities may be indicative of a static deformation of $^{32}$Mg. 
On the other hand, mean-field calculations \cite{RDN99} as well as  
quasiparticle random-phase approximation (QRPA) \cite{YG04} 
with the Skyrme interaction 
suggest that $^{32}$Mg may be spherical. In fact the energy ratio
between the first 4$^+$ and the first 2$^+$ states, $E_{4_1^+}/E_{2_1^+}$, is 
2.6 \cite{YSG01}, which is between the vibrational and rotational
limits \cite{YG04}. 

Note that the distorted-wave Born approximation (DWBA) yields 
identical results for both
rotational and vibrational couplings (to first order). 
In order to discriminate whether the transitions are vibration-like
or rotation-like, at least second-step processes (reorientation
and/or couplings to higher members) are necessary. 
The coupling effect plays a more important role in low-energy
reactions than at high and intermediate energies. Therefore, 
quasi-elastic scattering around the Coulomb barrier may provide
a useful method of clarifying the nature of the quadrupole collectivity of
$^{32}$Mg. 

Fig. 9 shows the excitation function of the quasi-elastic scattering
(upper panel) and the quasi-elastic barrier distribution 
(lower panel) for this system. 
The solid and dashed lines are results of coupled-channels
calculations where $^{32}$Mg is assumed to be a rotational or
a vibrational nucleus, respectively. We estimate the 
coupling strength $\beta_2$
from the measured $B(E2)$ value \cite{MIA95} to be 0.51. 
We include the quadrupole excitations in $^{32}$Mg up to the second 
member (that is, the first 4$^+$ state in the rotational band for the
rotational coupling, or the double phonon state for the vibrational
coupling). In addition, we include the single octupole phonon
excitation at 2.615 MeV in $^{208}$Pb \cite{MBD99}. 
The potential parameters which we use are 
$V_0$=180 MeV, $r_0$=1.15 fm, and $a$=0.63 fm, 
that give the same barrier height ($B$=106.6
MeV) as the Aky\"uz-Winther potential \cite{AW81}. 
For the imaginary potential, we use 
$W$=50 MeV, $r_w$=1.0 fm, and $a_w$=0.4 fm, but the results are
insensitive to this as long as it is localized 
inside the barrier with a large enough strength. 
We use the computer code {\tt CQUEL} \cite{HR04} in order 
to integrate the
coupled-channels equations. This code is a version of 
{\tt CCFULL}\cite{HRK99}, 
where the coupling is treated to all orders in the coupling
hamiltonian and the isocentrifugal approximation is employed in order
to reduce the dimension of the coupled-channels equations. 
In the code {\tt CQUEL}, we use the regular boundary condition at the
origin, instead of the incoming boundary condition, and we remove the 
restriction of {\tt CCFULL}, which computes only the fusion cross
sections.

In the figure, we can see well separated peaks in the
quasi-elastic barrier distribution both for the rotational and for the
vibrational couplings. Moreover, the two lines are considerably
different at energies around and above the Coulomb barrier, although 
the two results are rather similar below the
barrier. 
We can thus expect that the
quasi-elastic barrier distribution can indeed be utilized 
to discriminate between the rotational and the
vibrational nature of the quadrupole collectivity in $^{32}$Mg, 
although these results might be somewhat perturbed by other
effects which are not considered in the present calculations, such as
double octupole-phonon excitations in the target, transfer
processes or hexadecapole deformations. 

\section{Summary}

The quasi-elastic barrier distribution is 
a counterpart of the fusion barrier distribution in the sense that the
former is related to the reflection probability of a potential barrier
while the latter is related to the transmission. In this paper, we
have studied some detailed properties of the quasi-elastic barrier
distribution. 
Using semi-classical perturbation theory, we have obtained
an analytic formula for the quasi-elastic barrier distribution for a 
single barrier (that is, the quasi-elastic test function). The formula 
indicates that this test function consists of two
factors: one is related to the effect of the nuclear distortion of
the classical trajectory, while the other is the reflection probability of
the potential barrier. Due to the nuclear distortion, we found
that the quasi-elastic barrier distribution is slightly less well
behaved than 
the fusion barrier distribution. For instance, the peak position 
of the quasi-elastic barrier
distribution slightly deviates from the barrier height, and 
it has a low-energy tail. Nevertheless, the
quasi-elastic barrier distribution behaves rather similarly to
that for fusion on the whole, and both are sensitive to the same 
nuclear structure effects. 

In multi-channels systems, the validity of the barrier distribution 
relies on the isocentrifugal approximation, where 
the angular momentum of the relative motion in each
channel is replaced by the total angular momentum $J$. 
We have examined the applicability of this approximation for scattering
processes and have found that it works
well at least for backward angles, where such experiments are performed. 

The measurement of quasi-elastic barrier distributions is well suited
to future experiments with low-intensity exotic beams. 
To illustrate this fact, we have discussed as an example,
the effect of quadrupole
excitations in the neutron-rich $^{32}$Mg nucleus on quasi-elastic
scattering around the Coulomb barrier, and argued that the
quasi-elastic barrier distribution would provide a useful tool to 
clarify whether $^{32}$Mg is spherical or deformed. 
In this way, we expect that
the barrier distribution method will open up a 
novel means to allow 
the detailed study of the structure of neutron-rich nuclei in 
the near future. 

\section*{acknowledgments}

We thank E. Piasecki and E. Crema for helpful discussions. 

\begin{appendix}

\section{Semi-classcial perturbation theory}

In this Appendix, we derive Eq. (\ref{ratio}) for the backward-angle
elastic cross section using semi-classical perturbation theory. 
Our formula is an improvement of the one in Ref. \cite{LW81}, since we 
take into account the effect of 
nuclear distortion of the classical trajectory\cite{BS81}.

The scattering amplitude $f(\theta)$ for a spherical optical potential is 
given by, 
\begin{equation}
f(\theta)=\frac{1}{2ik}\sum_l(2l+1)P_l(\cos\theta)(S_l-1),
\label{f}
\end{equation}
where $\theta$ is the scattering angle and $k=\sqrt{2\mu E/\hbar^2}$. 
Since we are interested in backward scattering near $\theta \sim \pi$, 
we replace the Legendre polynomials $P_l(\cos\theta)$ with their 
asymptotic form, 
\begin{equation}
P_l(\cos\theta)\sim(-)^l\sqrt{\frac{\pi-\theta}{\sin\theta}}
J_0\left((l+\frac{1}{2})(\pi-\theta)\right),
\end{equation}
where $J_0(\theta)$ is the Bessel function. 
We now apply the well known Poisson sum formula to Eq. (\ref{f}) to obtain 
\begin{equation}
f(\theta)=\frac{1}{k}
\sqrt{\frac{\pi-\theta}{\sin\theta}}\sum_n(-)^n\int^\infty_0d\lambda\,
\lambda S(\lambda)J_0(\lambda(\pi-\theta))e^{(2n-1)i\pi\lambda},
\label{fback}
\end{equation}
where $\lambda=l+1/2$. 
At energies around the Coulomb barrier and for backward scattering, 
the contribution from $n=0$ dominates the sum in Eq. (\ref{fback})
\cite{B85}. Taking only $n=0$ and evaluating the integral in the
stationary phase approximation, one obtains (see Sec. 5.7 of
Ref. \cite{B85}), 
\begin{equation}
f(\theta)\sim\sqrt{\frac{\lambda}{k^2\sin\theta|\Theta'(\lambda)|}}
e^{-i(\lambda\theta-\pi/2)}S(\lambda),
\end{equation}
where 
$\Theta(\lambda)=2Re\,\delta'(\lambda)$ is the deflection function, 
$\delta(\lambda)$ being the phase shift, 
and $\lambda$ satisfies the stationary phase condition 
$\Theta(\lambda)=\theta$. Here, the dash denotes the derivative with
respect to the argument. This equation yields 
\begin{equation}
\frac{\sigma(\theta)}{\sigma_R(\theta)}
=\left|\frac{\Theta'_c(\lambda_c)}{\Theta'(\lambda)}\right|\,
\frac{\lambda}{\lambda_c}|S(\lambda)|^2. 
\label{ratio-sc}
\end{equation}

Landowne and Wolter evaluated Eq. (\ref{ratio-sc}) using a perturbation 
theory based on the semiclassical approximation \cite{LW81}. 
The stationary condition $\Theta(\lambda)=\Theta_c(\lambda_c)=\theta$ 
and the definition of the nuclear deflection function, 
$\Theta(\lambda)=\Theta_c(\lambda)+\Theta_N(\lambda)$, yield \cite{LW81}
\begin{equation}
\left|\frac{\Theta'_c(\lambda_c)}{\Theta'(\lambda)}\right|\,
\frac{\lambda}{\lambda_c}\sim 1+\frac{\eta}{2\lambda_c}\Theta_N(\lambda_c)
+\frac{\eta}{2}\Theta'_N(\lambda_c),
\label{scperturbation}
\end{equation}
to first order in $\lambda-\lambda_c$. In deriving this equation, 
we have assumed that $\eta$ is much larger than $\lambda_c$. 
In the semiclassical approximation, the nuclear phase shift is 
given by \cite{B85} 
\begin{eqnarray}
\delta_N(\lambda)&=&\int^\infty_{r_1}k(r)dr-\int^\infty_{r_c}k_c(r)dr, 
\label{delta} \\
k(r)&=&\sqrt{2\mu(E-V_N(r)-V_C(r)-V_\lambda(r))/\hbar^2}, \\
k_c(r)&=&\sqrt{2\mu(E-V_C(r)-V_\lambda(r))/\hbar^2}, 
\end{eqnarray}
where 
$V_N(r)$ and 
$V_C(r)$ are the nuclear and the Coulomb potentials, respectively, and 
$V_\lambda(r)=\lambda^2\hbar^2/2\mu r^2$ is the centrifugal potential. 
The classical turning points $r_1$ and $r_c$ satisfy 
$k(r_1)=k_c(r_c)=0$. 
To first order in the nuclear potential, the semi-classical phase shift is 
given by 
\begin{equation}
\delta_N(\lambda)\sim
-\frac{\mu}{\hbar^2}\int^\infty_{r_c}
\frac{V_N(r)}{k_c(r)}dr. 
\end{equation}
Expanding $k_c(r)$ around $r=r_c$ and assuming that 
$V_N(r)\sim -V_0\,e^{-r/a}$ 
near $r_c$, one obtains \cite{B85,LW81}
\begin{equation}
2\delta_N(\lambda)\sim -V_N(r_c)\frac{\sqrt{2a\pi k\eta}}{E} 
+ O(\lambda^2/\eta^2).  
\label{phase0}
\end{equation}
Using the perturbative phase shift (\ref{phase0}) in Eq. 
(\ref{scperturbation}), 
Landowne and Wolter obtained a simple form for the backward 
cross sections which is given by\cite{LW81}, 
\begin{equation}
\frac{\sigma_{\rm el}(E,\theta)}{\sigma_R(E,\theta)}
=\left(1+\frac{V_N(r_c)}{ka}\,
\frac{\sqrt{2a\pi k\eta}}{E}\right)
\cdot |S(E,\lambda_c)|^2. 
\label{ratio-sc0}
\end{equation}

An improved formula may be obtained by taking into account 
the effect of nuclear distortion of the classical trajectory. 
To this end, we follow the method suggested by Brink and Satchler \cite{BS81}. 
Transforming the coordinate in the first integral in Eq. (\ref{delta}) 
to the one 
which satisfies $k(r)=k_c(s)$, the semi-classical phase shift may be 
expressed as 
\begin{equation}
\delta_N(\lambda)=\int^\infty_{r_c}k_c(s)\frac{d}{ds}(r(s)-s)ds 
=-\int^\infty_{r_c}(r(s)-s)\frac{d}{ds}k_c(s)\,ds.
\end{equation}
The condition $k(r)=k_c(s)$ yields
$0=V_N(s)+(V_N'(s)+V_C'(s)+V'_\lambda(s))(r-s)$ to first order in 
$r-s$. We thus obtain
\begin{eqnarray}
\delta_N(\lambda)&\sim&-\frac{\mu}{\hbar^2}\int^\infty_{r_c}
\frac{V_C'(s)+V_\lambda'(s)}{V_N'(s)+V_C'(s)+V_\lambda'(s)}
\cdot\frac{V_N(s)}{k_c(s)}ds, \label{brink}\\
&\sim&
\left[
1-\frac{V_N'(r_c)}{V_C'(r_c)+V_\lambda'(r_c)}\right]
\cdot \frac{\mu}{\hbar^2}\int^\infty_{r_c}
\frac{-V_N(r)}{k_c(r)}dr, \\
&\sim&
\left[
1-\frac{V_N'(r_c)}{V_C'(r_c)+V_\lambda'(r_c)}\right]
\cdot \left(\frac{-V_N(r_c)}{2}\right)\frac{\sqrt{2a\pi k\eta}}{E} 
+ O(\lambda^2/\eta^2).  
\label{phase}
\end{eqnarray}
Here, we have expanded $r-s$ with respect to $V_N$ in Eq. (\ref{brink}) 
and evaluated it at the radius $r_c$. 
Substituting Eq. (\ref{phase}) into Eq. (\ref{ratio-sc}), we finally 
obtain Eq. (\ref{ratio}). 

Fig. 10 compares the semi-classical formula with the exact result 
(solid line) for the 
$^{16}$O + $^{144}$Sm reaction. We use the same optical potential as 
in Sec. II. The dotted line is obtained by the semi-classical perturbation 
of Landowne and Wolter, Eqs. (\ref{phase0}) and (\ref{ratio-sc0}). 
The dashed line is the result of semi-classical approximation which takes 
into account the nuclear distortion, Eqs. (\ref{phase}) and (\ref{ratio}). 
We see that the semi-classical perturbation theory works reasonably well 
around the Coulomb barrier 
when the effect of nuclear distortion is included. 
The deviation of the nuclear phase shift from the exact solution 
above the barrier would be improved by using the full semi-classical 
phase shift \cite{BT77}. However, we note that 
the backward cross sections are already reproduced reasonably well 
even with the present semi-classical perturbation theory. 

\end{appendix}

\begin{figure}
  \begin{center}
    \leavevmode
    \parbox{0.9\textwidth}
           {\psfig{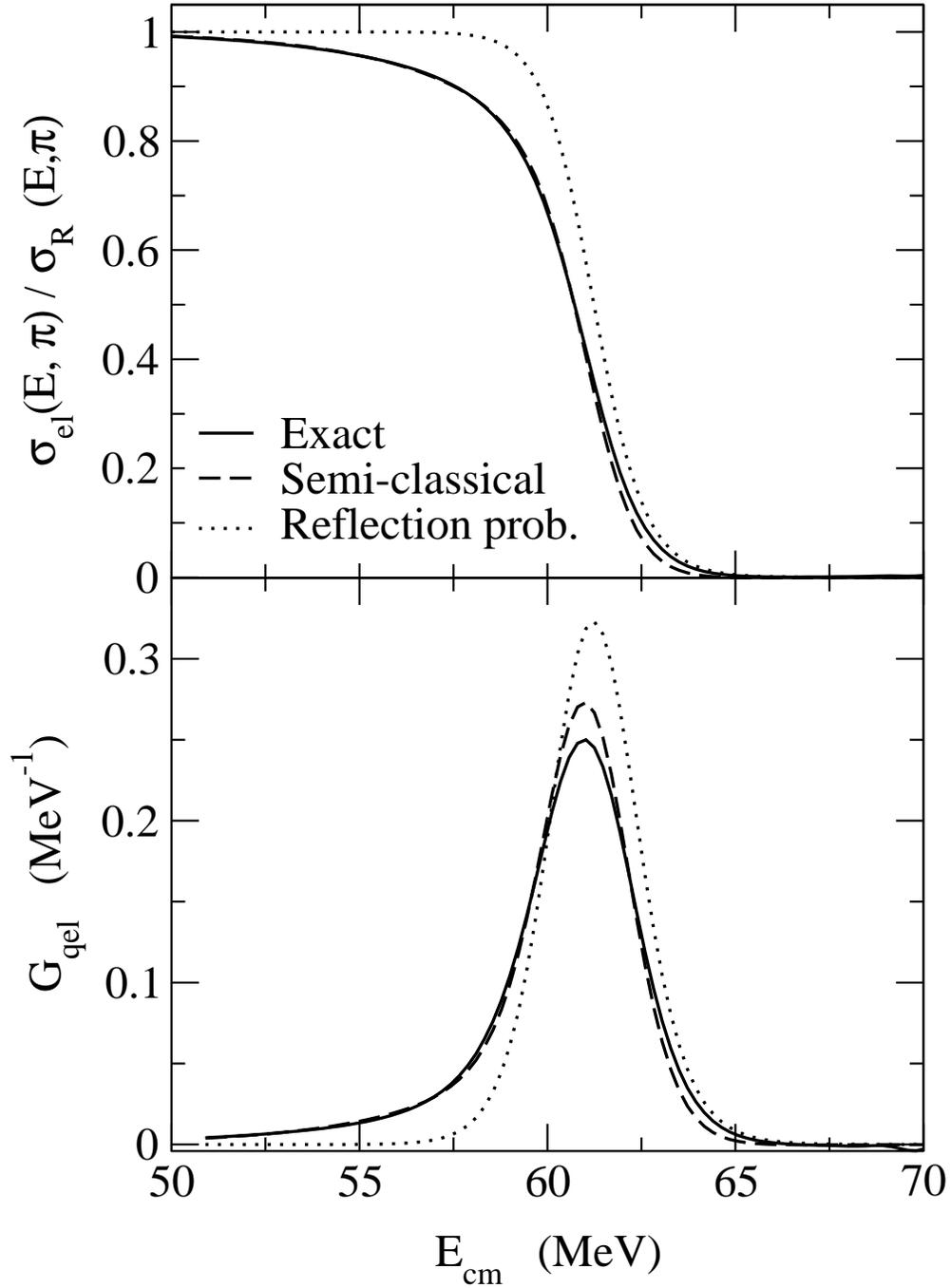}}
  \end{center}
\protect\caption{
The ratio of elastic scattering to the Rutherford cross section at 
$\theta=\pi$ (upper panel) and the quasi-elastic test
function $G_{\rm qel}(E)=-d/dE (\sigma_{\rm el}/\sigma_R)$ (lower panel) 
for the $^{16}$O + $^{144}$Sm reaction. 
The solid line is the exact solution of the optical potential, while
the dashed line is obtained with the semi-classical perturbation
theory. The dotted line denotes the reflection probability
$R(E)=|S(E)|^2$ for $s$-wave scattering.}
\end{figure}

\begin{figure}
  \begin{center}
    \leavevmode
    \parbox{0.9\textwidth}
           {\psfig{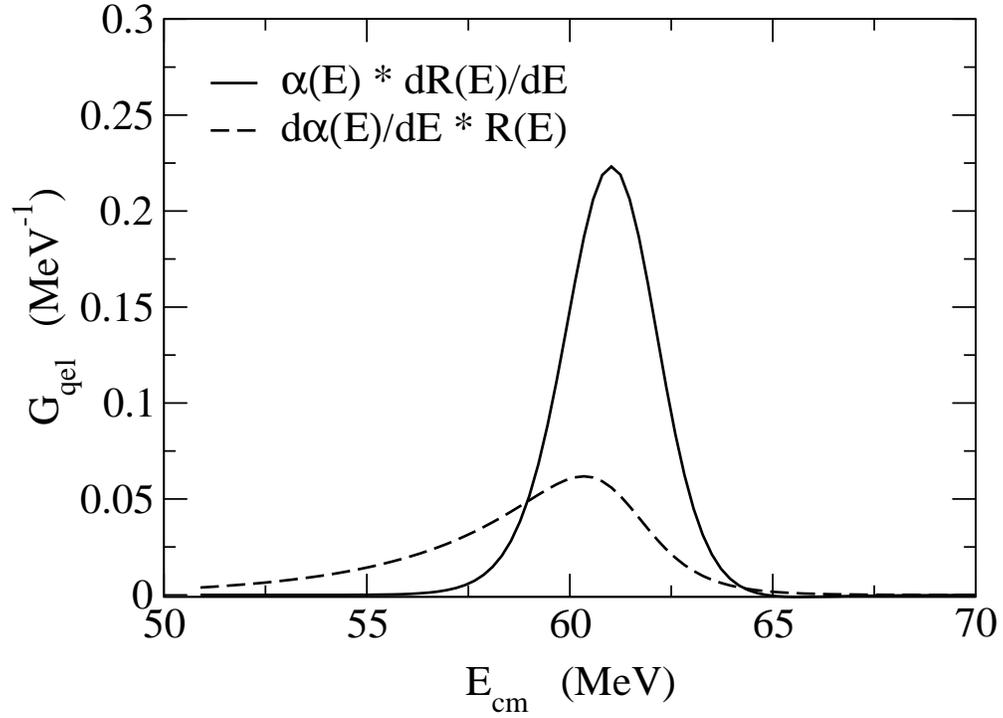}}
  \end{center}
\protect\caption{
Two separate contributions to the quasi-elastic test function. 
The solid line shows 
the function $\alpha(E)\cdot dR(E)/dE$, 
while the dashed line shows
$d\alpha(E)/dE\cdot R(E)$, where $\alpha(E)$ and $R(E)$ are the 
nuclear distortion function and the reflection probability, 
respectively. }
\end{figure}

\begin{figure}
  \begin{center}
    \leavevmode
    \parbox{0.9\textwidth}
           {\psfig{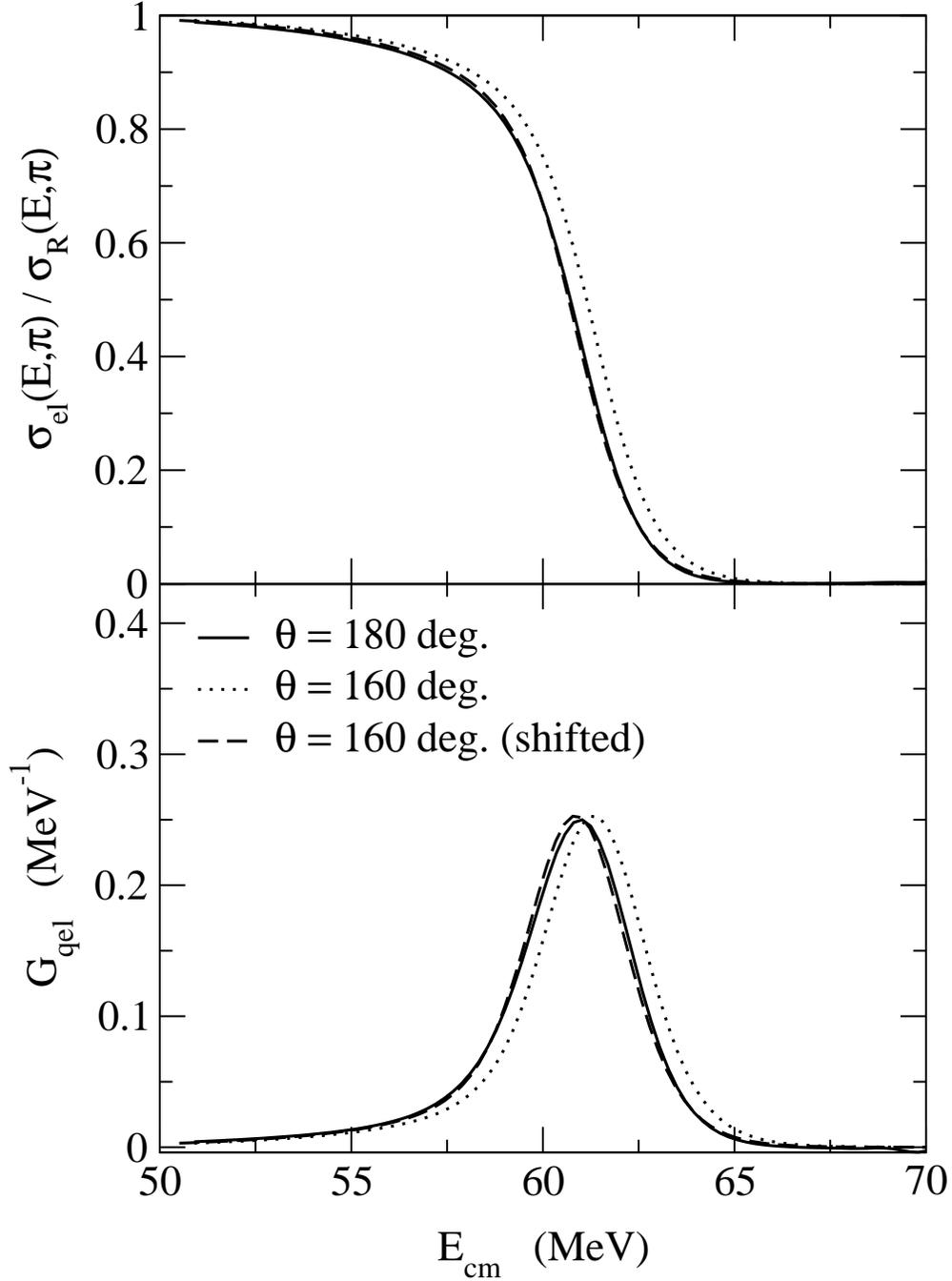}}
  \end{center}
\protect\caption{
Comparison of the 
ratio $\sigma_{\rm el}/\sigma_R$ (upper panel) and its energy 
derivative $-d/dE (\sigma_{\rm el}/\sigma_R)$ (lower panel) 
evaluated at two 
different scattering angles. The solid line is for $\theta=\pi$, while 
the dotted line is for $\theta=160^{\rm o}$. 
The dashed line is the same as the dotted line, but is shifted in energy 
by an amount equal to 
the centrifugal potential evaluated at the distance of closest
approach of the 
Rutherford trajectory. }
\end{figure}

\begin{figure}
  \begin{center}
    \leavevmode
    \parbox{0.9\textwidth}
           {\psfig{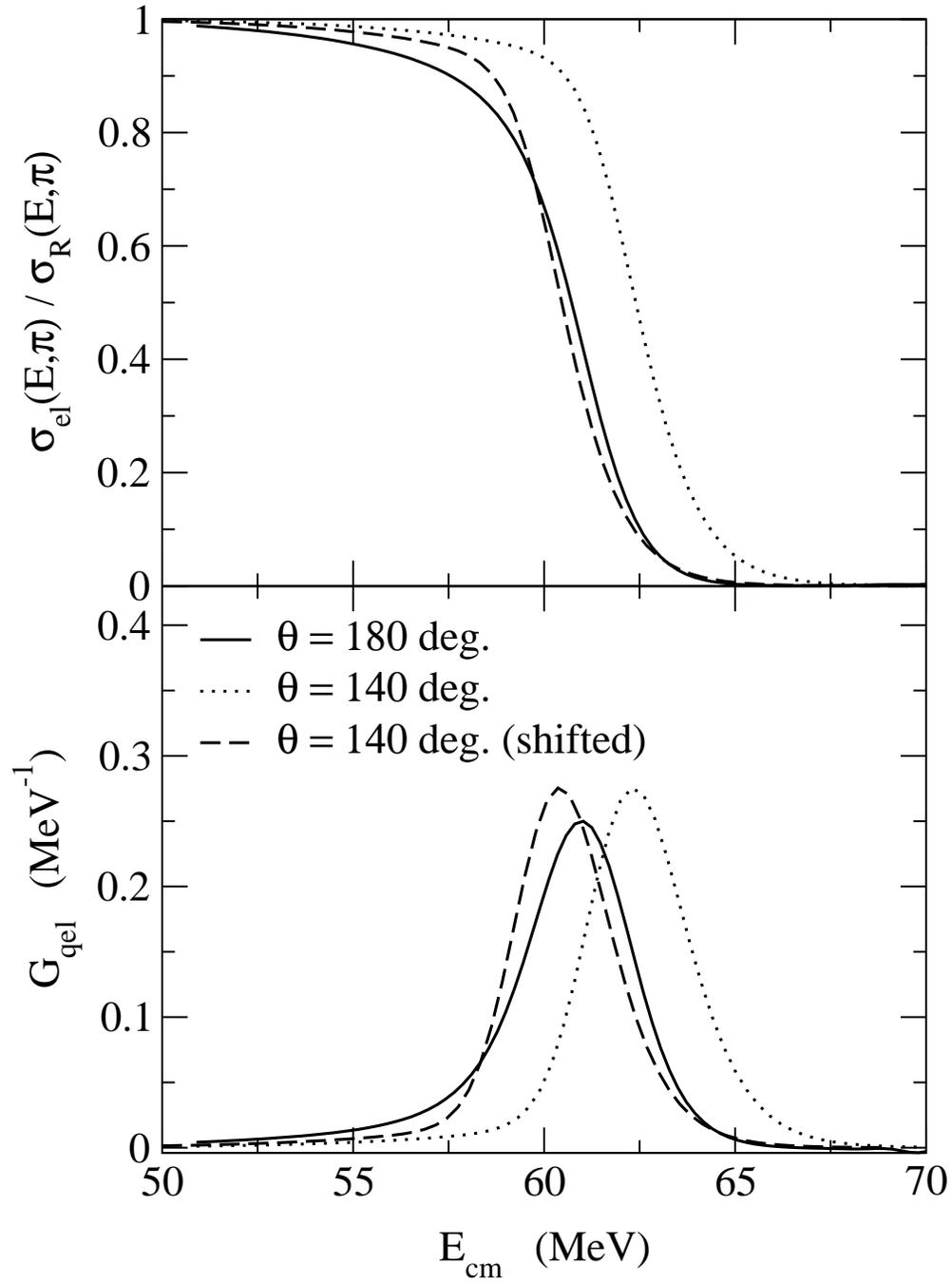}}
  \end{center}
\protect\caption{
The same as fig.3, but for 
$\theta=140^{\rm o}$. }
\end{figure}

\begin{figure}
  \begin{center}
    \leavevmode
    \parbox{0.9\textwidth}
           {\psfig{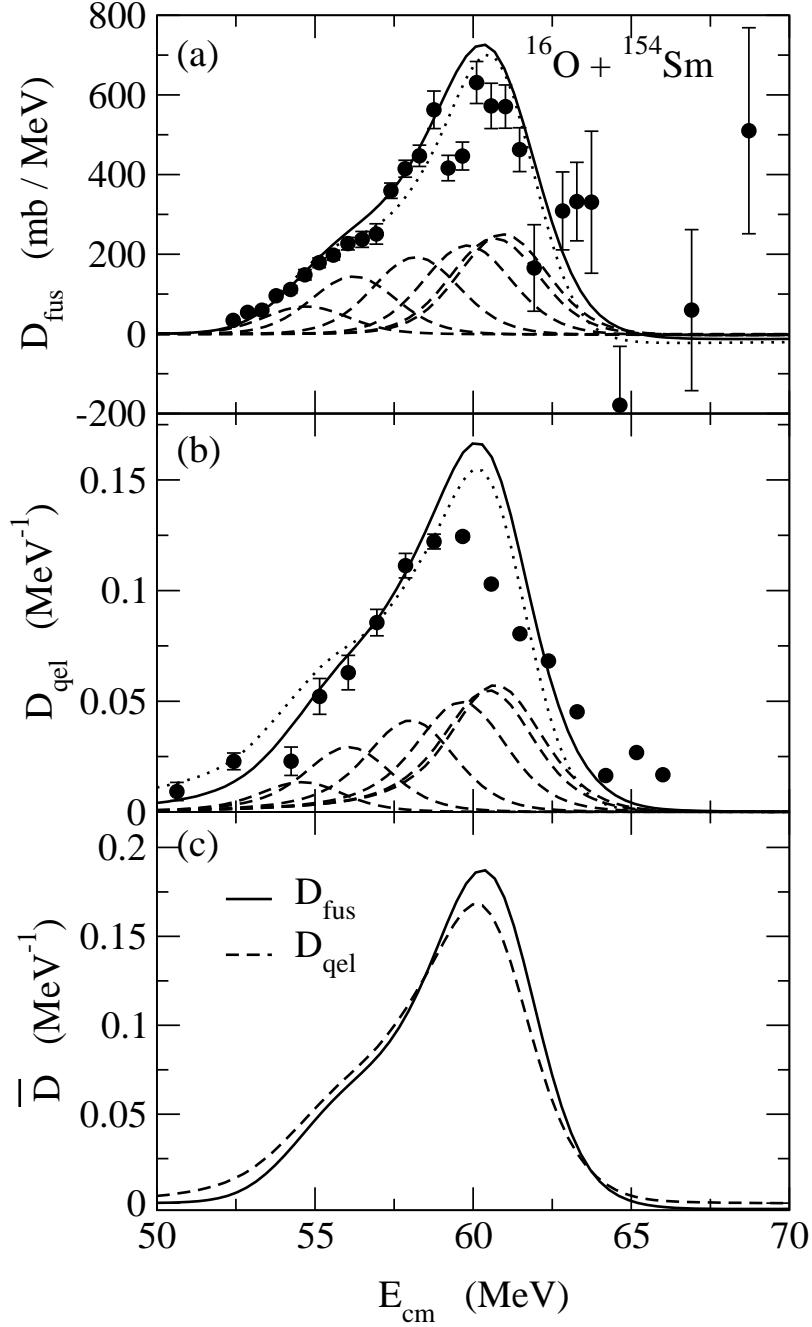}}
  \end{center}
\protect\caption{
(a) The fusion barrier distribution 
$D_{\rm fus}(E)=d^2(E\sigma_{\rm fus})/dE^2$ for the 
$^{16}$O + $^{154}$Sm reaction. The solid line is obtained with the
    orientation-integrated formula with $\beta_2=0.306$ and $\beta_4$=
    0.05. The dashed lines indicate the contributions from the six
    individual eigenbarriers. These lines are obtained by using a
Woods-Saxon potential with a surface diffuseness parameter $a$ of 0.65
fm. The dotted line is the fusion barrier distribution calculated with 
a potential which has $a$ = 1.05 fm. 
Experimental data are taken from
    Ref. {\protect\cite{L95}}. 
(b) Same as Fig. 5(a), but for the quasi-elastic barrier
    distribution 
$D_{\rm qel}(E)=-d[\sigma_{\rm qel}(E,\pi)/\sigma_R(E,\pi)]/dE$. 
Experimental data are from Ref. {\protect\cite{TLD95}}. 
(c) Comparison between 
the barrier distribution for fusion (solid line)
    and that for quasi-elastic scattering (dashed line). 
These functions are both normalized to unit area in the energy interval 
between 50 and 70 MeV.}
\end{figure}

\begin{figure}
  \begin{center}
    \leavevmode
    \parbox{0.9\textwidth}
           {\psfig{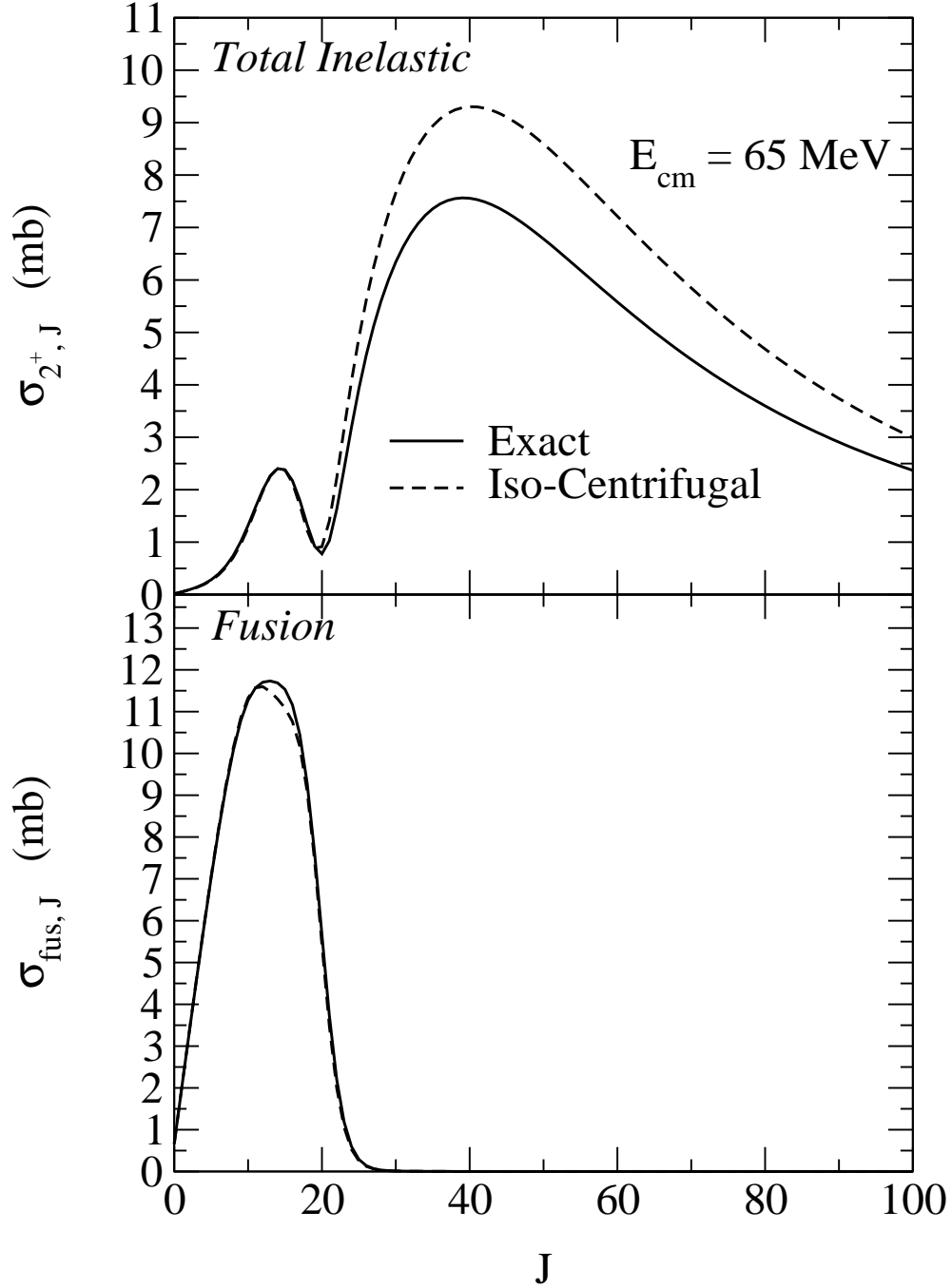}}
  \end{center}
\protect\caption{
The effect of a quadrupole-phonon excitation in the target nucleus on 
the partial cross sections for the $^{16}$O + $^{144}$Sm reaction at
$E_{\rm cm}$=65 MeV. 
The upper and the lower panels show 
the angle-integrated inelastic scattering
and the fusion cross sections, respectively. 
The solid line is the solution of 
the coupled-channels equations with the full angular momentum
coupling, while the dashed line is obtained in the isocentrifugal 
approximation. }
\end{figure}

\begin{figure}
  \begin{center}
    \leavevmode
    \parbox{0.9\textwidth}
           {\psfig{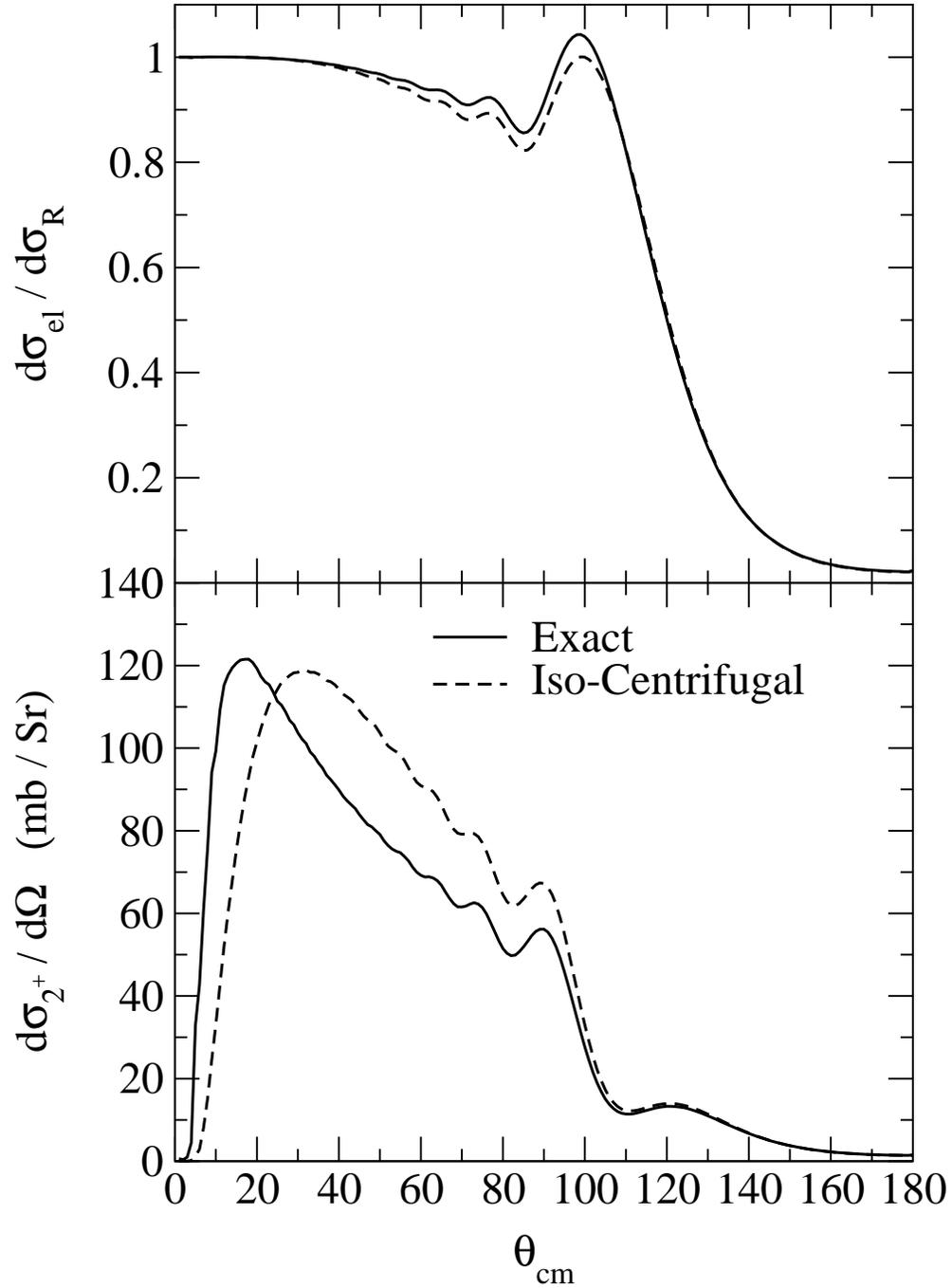}}
  \end{center}
\protect\caption{
The angular distributions for the elastic (upper panel) and the
    inelastic (lower panel) scattering for the $^{16}$O + $^{144}$Sm
    reaction at $E_{cm}$=65 MeV. 
The significance of each line is the same as in Fig. 5. }
\end{figure}

\begin{figure}
  \begin{center}
    \leavevmode
    \parbox{0.9\textwidth}
           {\psfig{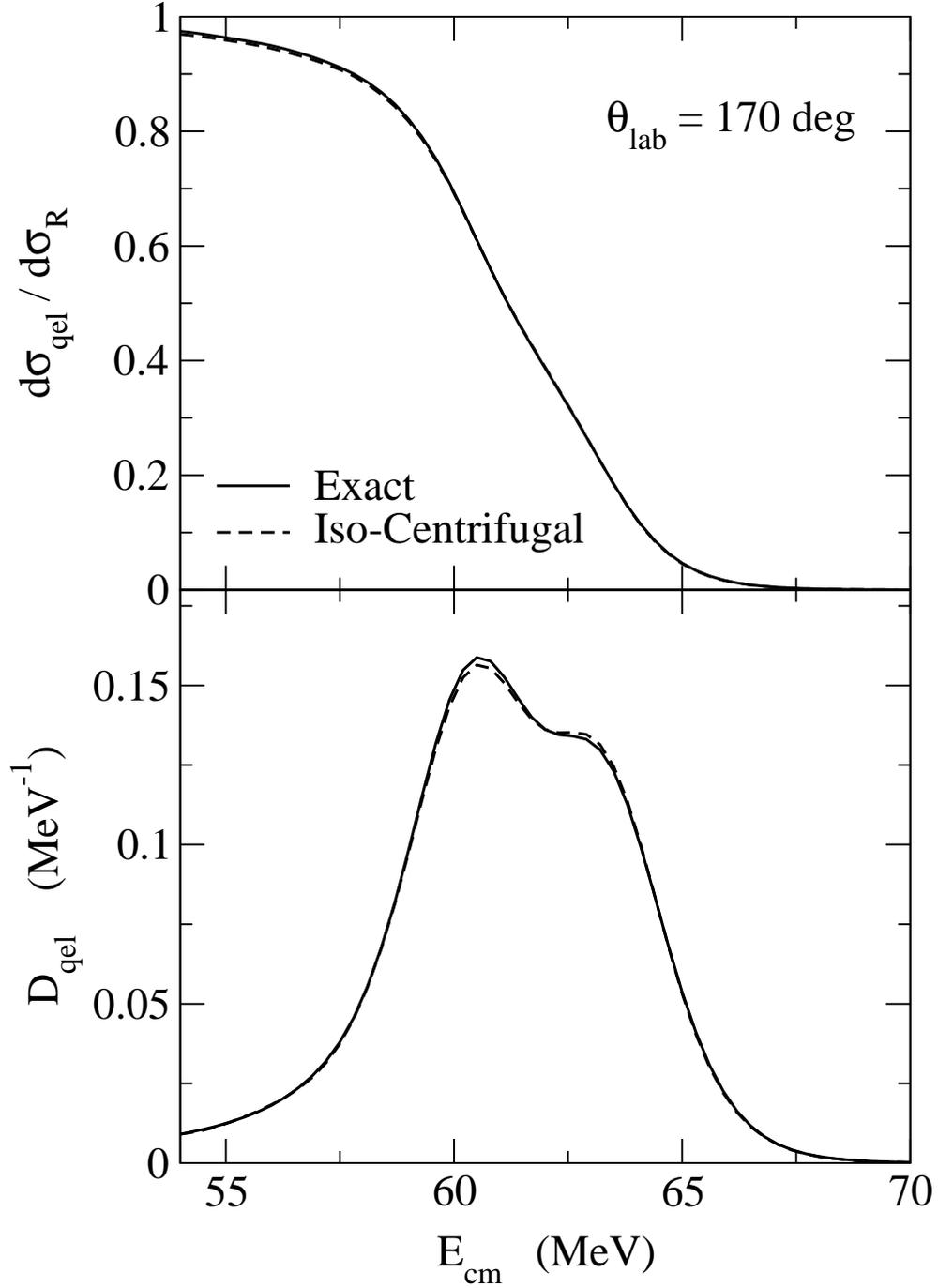}}
  \end{center}
\protect\caption{
The excitation function for quasi-elastic scattering (upper
    panel) and the quasi-elastic barrier distribution (lower
    panel) for the 
$^{16}$O + $^{144}$Sm reaction 
calculated at $\theta= 170^{\rm o}$ in the laboratory
    frame. 
The significance of each line is the same as in Fig. 5. }
\end{figure}

\begin{figure}
  \begin{center}
    \leavevmode
    \parbox{0.9\textwidth}
           {\psfig{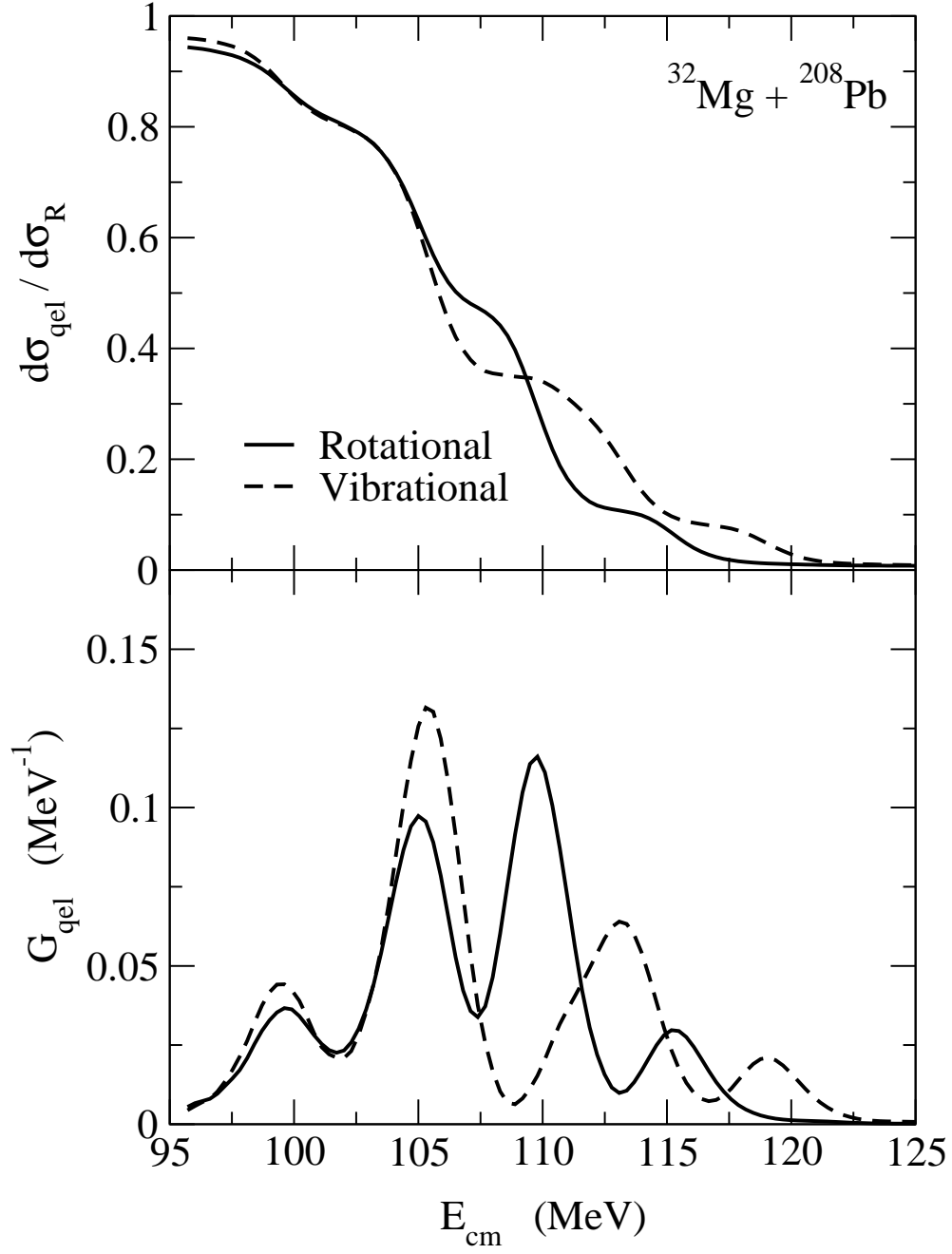}}
  \end{center}
\protect\caption{
The excitation function for quasi-elastic scattering  
(upper panel) and the quasi-elastic barrier distribution 
(lower panel) for the 
$^{32}$Mg + $^{208}$Pb reaction around the Coulomb barrier. 
The solid and the dashed lines are the results of coupled-channels
calculations which assume that $^{32}$Mg is a rotational and a
vibrational nucleus, respectively. 
The single octupole-phonon excitation in 
$^{208}$Pb is also included in the calculations. }
\end{figure}

\begin{figure}
  \begin{center}
    \leavevmode
    \parbox{0.9\textwidth}
           {\psfig{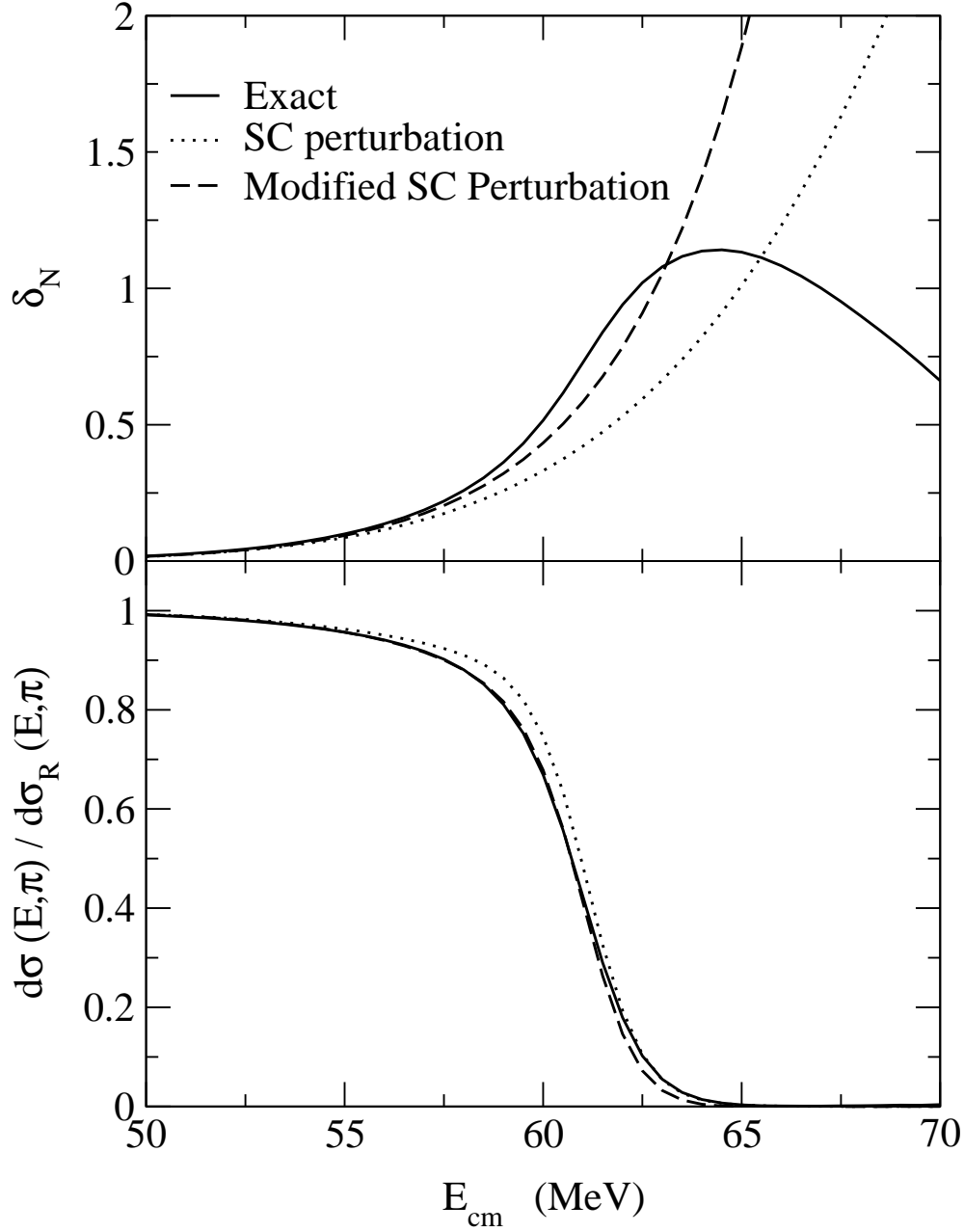}}
  \end{center}
\protect\caption{Comparison of the semi-classical formulae with the 
exact solution for the $^{16}$O+$^{144}$Sm reaction. The upper 
and the lower panels show the nuclear phase shift and the ratio of the 
elastic to the Rutherford cross sections at the scattering angle $\pi$, 
respectively. The solid line is obtained by numerically integrating 
the Schr\"odinger equation, while the dotted line is the result of 
the primitive semi-classical perturbation theory, 
Eqs. (\ref{phase0}) and (\ref{ratio-sc0}). 
The dashed line indicates the 
result of the semi-classical perturbation theory which takes 
into account the effect of nuclear distortion of the classical trajectory, 
Eqs. (\ref{phase}) and (\ref{ratio}). 
}
\end{figure}

\end{document}